# Representing Digital Assets using MPEG-21 Digital Item Declaration

Jeroen Bekaert[1,2], Herbert Van de Sompel[1]

[1] *Digital Library Research & Prototyping Team*
*Los Alamos National Laboratory*
*MS P362, PO Box 1663*
*Los Alamos, NM 87544-7113, USA*
{jbekaert, herbertv}@lanl.gov

[2] *Dept. of Architecture and Urbanism*
*Faculty of Engineering, Ghent University*
*Jozef-Plateaustraat 22, 9000 Gent, Belgium*
{jeroen.bekaert}@ugent.be

**Abstract** – Various XML-based approaches aimed at representing compound digital assets have emerged over the last several years. Approaches that are of specific relevance to the digital library community include the Metadata Encoding and Transmission Standard (METS), the IMS Content Packaging XML Binding, and the XML Formatted Data Units (XFDU) developed by CCSDS Panel 2. The MPEG-21 Digital Item Declaration (MPEG-21 DID) is another standard specifying the representation of digital assets in XML that, so far, has received little attention in the digital library community. This article gives a brief insight into the MPEG-21 standardization effort, highlights the major characteristics of the MPEG-21 DID Abstract Model, and describes the MPEG-21 Digital Item Declaration Language (MPEG-21 DIDL), an XML syntax for the representation of digital assets based on the MPEG-21 DID Abstract Model. Also, it briefly demonstrates the potential relevance of MPEG-21 DID to the digital library community by describing its use in the aDORe repository environment at the Research Library of the Los Alamos National Laboratory (LANL) for the representation of digital assets.

## 1. Introduction

Assets stored in digital libraries are typically more complex than autonomous files that can be characterized by a single MIME type. They typically aggregate multiple datastreams of a variety of MIME types. They also hold secondary information about those constituent datastreams, including information supporting discovery, digital preservation and rights management. Assets may be related to others, for example, by part-whole, sequential, and version relationships. And, contemporary digital library architectures also entertain the notion of associating dissemination capabilities with assets and parts thereof.

The ISO Open Archival Information System (OAIS) Reference Model refers [25] to these assets as OAIS Information Objects; the seminal Kahn/Wilensky framework [36] refers to them as Digital Objects. Indeed, Kahn & Wilensky define a Digital Object as more than just a set of bits; they regard a Digital Object as a data structure with a unique persistent identifier that, apart from accommodating constituent datastream(s), holds secondary information about those datastream(s).

Various techniques to represent digital assets have emerged from different communities over the last years, and a historical overview is presented in [46]. Of special interest in the current technological environment are techniques that represent a digital asset as an XML document. Examples include the Metadata Encoding and Transmission Standard (METS) [40], the IMS Content Packaging XML Binding [23], and the XML Formatted Data Units (XFDU) developed by CCSDS Panel 2 [17].

The second Part of MPEG-21, Digital Item Declaration (MPEG-21 DID) [28,18], is another technique that, so far, has received little attention in the digital library community. This article gives a brief overview of the MPEG-21 standardization effort, explores the MPEG-21 Digital Item Declaration (MPEG-21 DID), and gives examples of its use in a digital library context. The paper also introduces several other parts of MPEG-21, such as the MPEG-21 Digital Item Identification (MPEG-21 DII) [29], and discusses their integration with the MPEG-21 DID.

## 2. From MPEG-1 to MPEG-21

The Moving Picture Experts Group (MPEG) is a working group of ISO/IEC in charge of the development of standards for coded representation of digital audio and video. So far, several MPEG standards have had a significant impact on the multimedia landscape.

- The MPEG-1 (1993) and MPEG-2 (1996) standards have enabled the production of widely adopted commercial products, such as Video CD, MP3, digital audio broadcasting (DAB), DVD, and digital television (DVB and ATSC) [14,37].
- MPEG-4 standardized audiovisual coding solutions that address the needs of communication, interactive and broadcasting services. MPEG-4 is currently used on the Internet (e.g. as QuickTime 6) [48].
- More recently, MPEG-7, formally named 'Multimedia Content Description Interface', concentrated on the description of multimedia content. While previous MPEG standards have focused on the coded representation of audio-visual content, MPEG-7 is primarily concerned with secondary textual information about the audio-visual content [22,26].

Together, the MPEG-1, -2, -4, and -7 standards provide a powerful set of specifications for multimedia representation. Many other high-quality multimedia standards (and proprietary solutions), such as JPEG and JPEG 2000, are being created to meet the needs of different communities. However, widespread deployment of interoperable multimedia applications requires more than just this array of standards that are mainly oriented towards file based media types. Standards are needed that focus on the representation of content consisting of multiple files; standards that specify how applications can interact with such content in an interoperable way; standards that facilitate the transfer, adaptation and dissemination of content; standards that allow for the identification and promote the protection of the content and the rights of the rights holders, and so on.

MPEG-21, formally called the 'Multimedia Framework', seeks to fill these gaps. Its vision is *'to define a normative set of tools for multimedia delivery and consumption for use by all the players in the delivery and consumption chain'* [27,42]. In order to facilitate interoperability within a domain or between domains, those tools may be used independently or selectively in combination. The envisioned techniques endeavor to cover the entire content delivery chain, encompassing content creation, protection, adaptation, dissemination and consumption.

## 3. MPEG-21 basic concepts

MPEG-21 introduces the Digital Item as a *'structured digital object with a standard representation, identification and metadata'* [27,42]. The Digital Item is the digital representation of an asset. It is the unit that is acted upon within the MPEG-21 framework. The Digital Item of MPEG-21 can roughly be considered the equivalent of the Digital Object of the Kahn/Wilensky framework.

Parties that interact within the MPEG-21 environment are categorized as Users. The User roles include creators, consumers, rights holders, content providers, distributors, and so on. There is no technical distinction between providers and consumers. All Users interact with Digital Items.

The goal of MPEG-21 can thus be regarded as to provide a set of tools which enables a User to interact with another User and the object of that interaction is a Digital Item. Interactions include providing content, modifying content, archiving content, delivering content, consuming content, subscribing to content, etc.

# 4. Parts of MPEG-21

MPEG-21 is organized into several independent parts (currently eighteen) primarily to allow various slices of the technology to be used autonomously. Although the various parts can be used separately, they were developed to give optimal results when used together. The most important MPEG-21 parts for the purpose of this paper are:

- MPEG-21 Part 2 – Digital Item Declaration (henceforth referred to as MPEG-21 DID), detailing the representation of Digital Items [28,18].
- MPEG-21 Part 3 – Digital Item Identification (henceforth referred to as MPEG-21 DII), detailing the identification of Digital Items and their contained entities [29].
- MPEG-21 Part 4 – Intellectual Property Management and Protection (henceforth referred to as MPEG-21 IPMP), detailing a framework to enforce licenses, expressed using MPEG-21 REL [52].
- MPEG-21 Part 5 – Rights Expression Language (henceforth referred to as MPEG-21 REL), detailing a language to express rights pertaining to Digital Items and/or parts thereof [30].
- MPEG-21 Part 6 – Rights Data Dictionary (henceforth referred to as MPEG-21 RDD), detailing a set of clear, structured, and uniquely identified terms that can be used to support rights expression languages, such as the MPEG-21 REL [31].
- MPEG-21 Part 7 – Digital Item Adaptation (henceforth referred to as MPEG-21 DIA), detailing the adaptation and transcoding of datastreams based on contextual information such as device capabilities, network characteristics and User preferences [32].
- MPEG-21 Part 10 – Digital Item Processing (henceforth referred to as MPEG-21 DIP), detailing the association of processing information with Digital Items and/or parts thereof [20].

Although MPEG-21 originates in a community that focuses on the coded representation of audio and video, there is a clear overlap between the problem domain addressed by the MPEG-21 effort, and ongoing efforts regarding the interoperable representation, management, and dissemination of digital assets in the digital library community. For example, MPEG-21 DID and MPEG-21 DII directly relate to the aforementioned XML-based representational approaches, such as METS and IMS. Also, MPEG-21 DIP and MPEG-21 DIA reveal a remarkable parallel with sophisticated architectures that emerged from the digital library community, specifically FEDORA [47,49].

# 5. MPEG-21 DID: Digital Item Declaration

In the MPEG-21 Framework, Digital Items are modeled according to the second Part of MPEG-21: the Digital Item Declaration (MPEG-21 DID) standard [28,18]. MPEG-21 DID specifies the declaration of Digital Items in three distinct sections:

- *Abstract Model*: A set of abstract terms and concepts that, together, form a well-defined data model for declaring Digital Items. The Abstract Model allows for Digital Items to be represented in many ways. So far, MPEG-21 Part 2 defines a normative XML representation of Digital Items based on the Model. But, the Model allows for the emergence of an RDF instantiation, a non-XML-based binary instantiation, etc. The declaration of a Digital Item compliant with the Abstract Model is referred to as a Digital Item Declaration (DID). See also Figure 1 and Section 5.1.
- *Representation of the Model in XML*: The description of an XML syntax for each of the entities defined in the Abstract Model. This XML-based syntax is referred to as the MPEG-21 Digital Item Declaration Language (MPEG-21 DIDL). A DID represented according to the MPEG-21 DIDL syntax is referred to as a DIDL document. See also Figure 1 and Section 5.2.
- *XML Schema*: The W3C XML Schema's [21] specifying the MPEG-21 DIDL syntax and constrains for the structure of DIDL documents.

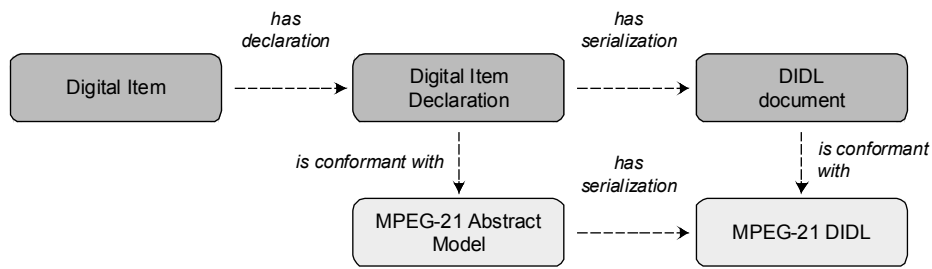

**Figure 1**: Relationship between MPEG-21 Abstract Model and MPEG-21 DIDL

The first edition of MPEG-21 Part 2 was published as an ISO (Internation Organization for Standardization) standard in March 2003 [28]. A second edition [18] has been finalized mid 2005 and mainly enhances the functionality of the MPEG-21 DIDL. The lion's share of those enhancements are the result of amendments and suggestions proposed by the authors, and are driven by a digital library use case of MPEG-21 technologies. The differences between both editions will be emphasized in the text. Once approved for publication, the second edition of MPEG-21 DID will be freely available from the ISO website [http://www.iso.org].

## 5.1. Abstract Model

The introduction of a data model is a characteristic that distinguishes MPEG-21 DID from related techniques aimed at representing digital assets. The existence of a data model provides flexibility for deployment of compatible digital assets representations in various technical environments. The MPEG-21 Abstract Model defines several constituent entities of a DID. This section provides a simplified explanation of each of those entities. When a reference is made to entities of the Abstract Model, the *italic* font style is used. Interested readers are referred to [20] for full details of the Model.

A first set of entities consists of the core building blocks for declaring Digital Items. Some are shown in Figure 2. These entities make up the backbone of a DID and are presented in a bottom-to-top approach, starting at the leaf of the tree with the *resource* entity, representing an actual datastream of a digital asset:

- *resource*: A *resource* is an individually identifiable datastream such as a video file, image, audio clip, or a textual asset. A *resource* may also potentially be an identifiable non-digital thing.
- *component*: A *component* is the binding of one or more equivalent *resources* to a (set of) *descriptor/statement* construct(s). These *descriptor/statement* constructs(s) contain secondary information related to all the *resource* entities bound by the *component*. A *component* is considered a dummy entity that is merely used to group *resources* and secondary information conveyed in one or more associated *descriptor/statement* construct(s).
- *item* or Digital Item: An *item* is the binding of one or more *items* and/or *components* to a (set of) *descriptor/statement* construct(s). These *descriptor/statement* construct(s) contain information about the represented *item*. In the Abstract Model, an *item* entity is equivalent to a Digital Item. Hence, *items* are the first point of entry to the content for a User. If *items* contain other *items*, then, the outermost *item* represents the composite *item*, and the inner *items* represent the individual *items* that make up the composite.
- *container*: A *container* is the binding of one or more *items* and/or *containers* to a (set of) *descriptor/statement* construct(s). The *containers* can be used to form groupings of Digital Items. The *descriptor/statement* constructs contain information about the represented *container*.
- A *descriptor/statement* construct, as shown in Figure 2, introduces an extensible mechanism that can be used to associate information with other entities of the Abstract Model. A *descriptor/statement* construct typically associates textual secondary information with its enclosing entity. For example, a *descriptor/statement* construct attached as a child entity to an *item* provides secondary information about that *item*. Similarly, a *descriptor/statement* construct attached as a child entity to a *container* provides secondary information about that *container*. The usage of

*descriptor/statement* constructs attached to a *component* deviates from the default rule in that they provide information about the *resources* bound by the *component*, and not about the enclosing entity. Examples of likely secondary information (or '*statements*') include information supporting discovery, digital preservation and rights expressions. *Statements* are non-identifiable assets, and, as a result, cannot have licenses associated with them.

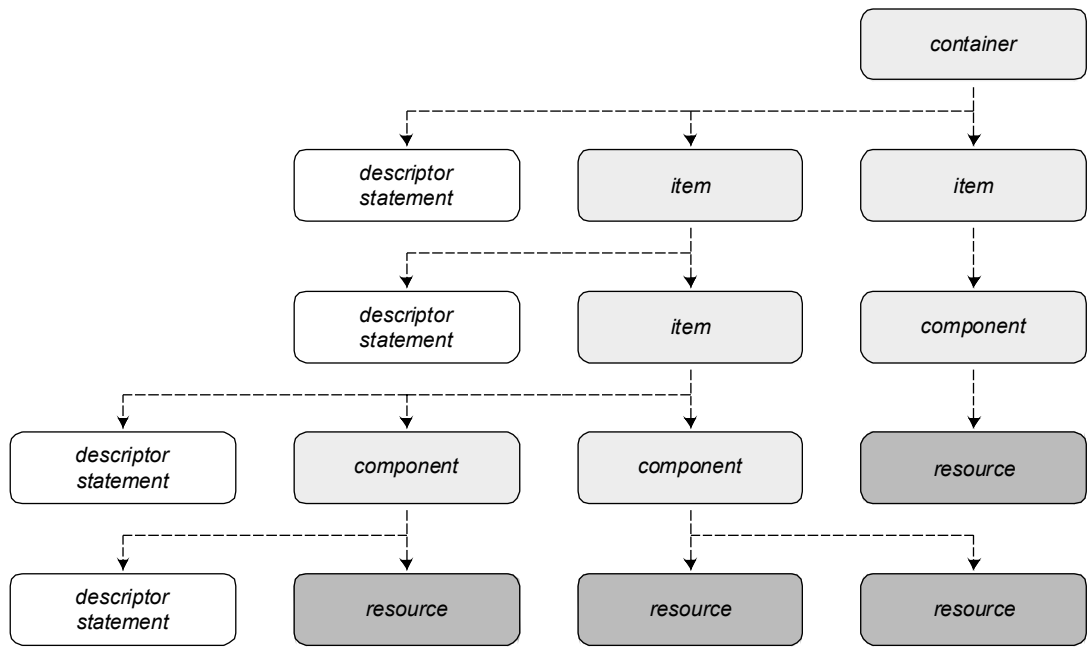

**Figure 2**: Example of a DID structure conformant with the MPEG-21 DID Abstract Model

A next set of entities, shown in Figure 3, refines the description of the *resource* entity:
- *fragment*: A *fragment* designates a specific point or range within a *resource*. *Fragments* may be *resource* type specific.
- *anchor*: An *anchor* binds one or more *descriptor/statement* constructs to a *fragment*. These *descriptor/statement* constructs contain information related to the *fragments* bound by the *anchor*. Similarly to the *component* entity, an *anchor* should be considered a dummy entity used merely to group *fragment* and *descriptor/statement* constructs.

For example, a *fragment* may specify a polygonal area within an image *resource* or a specific point in time of an audio track. The *anchor* entity binds secondary information to the polygonal area of the image or the time point of the audio track, respectively.

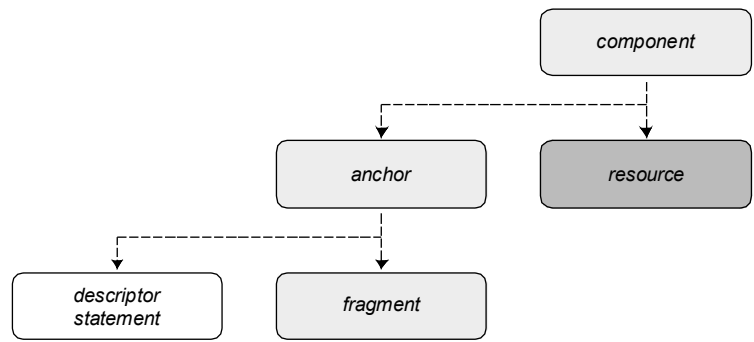

**Figure 3**: Example of a *component* grouping an *anchor* and *resource*.

Another entity of the DID Abstract Model is *annotation*. An *annotation* adds information to an identified entity of the model without altering or adding something to that entity. The initial source entity remains intact. The information conveyed by an *annotation*, can take the form of an *assertion*, *descriptor/statement*, or *anchor*.

A last set of entities, containing the *choice*, *selection*, *condition* and *assertion* entities, allows describing a Digital Item (or a part thereof) as being optional or available under specific conditions. For example, a *choice* in a DID may represent the choice between a high bandwidth internet connection and a dial-up connection. Based on the *selection* made by a User, the *conditions* attached to (an entity of) a Digital Item may be fulfilled and (the entity of) the Digital Item may become available. As such, dependent on the nature of the *conditions*, the entity could contain a high resolution datastream or a compressed file, respectively. The details of these conditional mechanisms are beyond the scope of this paper. Interested readers are referred to the MPEG-21 DID specification for full details.

## 5.2. Representation of the Abstract Model in XML

The Digital Item Declaration Language (MPEG-21 DIDL) is the normative XML Representation of the Abstract Model. The entities defined in the Abstract Model are each represented in MPEG-21 DIDL by a like-named XML element. In the remainder of this paper, the `courier` font is used to refer to these XML elements. For example, the *component* entity of the Abstract Model is represented in MPEG-21 DIDL by the `Component` XML element. The semantics and structural positions of the DIDL elements correspond with those of the entities of the Abstract Model. The MPEG-21 DIDL also defines some special elements that do not correspond to any of the Abstract Model entities. These elements can only exist in the XML Representation of the Abstract Model:

- The `DIDL` element is the root element of a DIDL XML document. The second edition of MPEG-21 DID allows for the association of secondary information pertaining to the DIDL document itself by adding attributes to the `DIDL` root element or by including XML elements in the `DIDLInfo` element that itself is a child of the `DIDL` root element. Also, the `DIDL` root element may have an optional `DIDLDocumentId` attribute. This attribute can be used to convey the identifier of the DIDL document. Identifiers of the digital asset(s) or Digital Item(s) declared within the DIDL document are not conveyed by this attribute, but rather by a `Descriptor/Statement` construct as explained in Section 6.3. Both the `DIDLInfo` element and the `DIDLDocumentId` attribute have resulted from proposals submitted by the authors and are largely inspired by a digital library and digital archive use case of MPEG-21 DID [6,10]. Indeed, there is a clear parallel between information that resides in the `DIDLInfo` element and the notion of OAIS [25] Packaging Information as used in the archiving domain.
- The `Reference` element represents a link to another XML element of a DIDL document, and virtually embeds the contents of the referenced element into the referring element. References can be made to elements within the same DIDL document or to elements in another DIDL document. The former type of reference is known as an internal reference; the latter is known as an external reference. An internal reference allows a single source to be maintained for an element that occurs in more than one place in a DIDL document. An external reference allows a DIDL document to be split up into multiple linked discrete DIDL documents. In the second edition of MPEG-21 DID, the `Reference` element has been removed as similar functionality can be achieved by using XML Inclusion 1.0 (XInclude) [41]. The latter has been published as a W3C Recommendation in December 2004.
- The `Declarations` element is used to define DIDL elements in a DIDL document without actually instantiating them. A declared element can then be instantiated by using a `Reference` element.

The grammar and syntax of the MPEG-21 DIDL is captured in a W3C XML Schema [21]. The XML Schema of the second edition of MPEG-21 DID is available from the ISO website [http://standards.iso.org/ittf/PubliclyAvailableStandards/MPEG-21_schema_files/]. To be conformant with the MPEG-21 DIDL syntax, validation against the grammar of the XML Schema is necessary, yet not sufficient. Some mechanisms, such as Schematron [33], may be required to express XPath [15] based rules that cannot be expressed by means of the XML Schema. The full specifics of these rules go beyond the scope of this paper. Interested readers are referred to [20] for full details of the MPEG-21 DIDL syntax.

# 6. A step by step example of a DIDL document

In this section, XML excerpts of a sample DIDL document are provided to illustrate the main characteristics of MPEG-21 DIDL in the context of its use in the digital library domain. The DIDL document is shown in full in Table 9. The document is the MPEG-21 DIDL-based XML representation of a simple digital asset consisting of:

- A scholarly publication with identifier 'info:doi/10.1045/july95-arms'. Two versions of the publication are provided as separate datastreams: the original PostScript file, and a derived PDF file.
- Descriptive secondary information about the publication expressed in the Dublin Core format [44].

## 6.1. Providing constituent datastreams

Each constituent datastream of a Digital Item is accommodated in a separate `Resource` element, representing a *resource*. As shown in Table 1, MPEG-21 DIDL allows those *resources* to be physically embedded within a DIDL document – i.e. 'By Value' provision – or to be pointed at from within a DIDL document – i.e. 'By Reference' provision. Table 1 also shows how the nature of those assets relates to the way in which they can be included By Value. MPEG-21 DIDL allows for the inclusion of unencoded (mixed) XML elements not defined by the MPEG-21 DIDL Schema. In addition, binary datastreams may be provided By Value using a base64 encoding as specified in IETF RFC 3548 [35]. The white rectangles in Table 1 indicate provision techniques that became available in the second edition of MPEG-21 DID as a result of amendments proposed by the authors [7,9].

|              |                      | Resource | Statement |
|--------------|----------------------|----------|-----------|
| **By Value** | (mixed) XML elements | ☐        | ■         |
|              | base64 encoded data  | ■        | ■         |
| **By Reference** | N/A              | ■        | ☐         |

**Table 1**: 'By Value' and 'By Reference' provision of assets and secondary information in a DIDL document

Table 2 demonstrates a 'By Reference' provision of both the PDF and Postscript constituents of the sample Digital Item. The `ref` attribute contains the actual network location. A 'By Value' provision of the PostScript and PDF assets can be accomplished by base64 encoding the binary data and wrapping the outcome in the `Resource` element. As a result of amendments proposed by the authors, the second edition of MPEG-21 DID requires the use of the `encoding` attribute (with a value set to 'base64'), whenever the `Resource` element contains base64 encoded data [7,9]. If the `encoding` attribute is omitted, the data must be un-encoded. This removes an ambiguity in the first edition of MPEG-21 DID, which does not support an unambiguous way to determine whether embedded character data is base64 encoded or not. Table 3 shows an abbreviated By Value provision of the PDF file. In addition to the `ref` and `encoding` attributes, a mandatory `mimeType` attribute indicates the MIME media and subtype of the provided datastreams.

```
<didl:Resource mimeType="application/pdf" ref="http://purl.lanl.gov/tech/pdf/015997845.pdf"/>
```

Table 2: A `Resource` element providing the PDF datastream 'By Reference'

```
<didl:Resource mimeType="application/pdf" encoding="base64">
JVBERi0xLjMKJeLjz9MNCjEgMCBvYmoKPDwgCi9TdWJqZWN0ICgpCi9LZXl3b3JkcyAoKQovQ3Jl
YXRvciAoWFBQKQovVGl0bGUgKCkKL1Byb2R1Y2VyICgpCi9Nb2REYXRlIChEOjIwMDQwNTE0MTIy
MzE2KQovQ3JlYXRpb25EYXRlICgyMDA0MDUxNDEyMjMwMSkKL0F1dGhvciAoKQo+PiAKZW5kb2Jq
CjIgMCBvYmoKPDwgCi9UeXBlIC9QYWdlIAovUGFyZW50IDExIDAgUiAKL1Jlc291cmNlcyA0IDAgUiAg...</didl:Resource>
```

Table 3: A `Resource` element providing the PDF datastream file 'By Value'

It may be important to note that the second edition of MPEG-21 DID also allows for expressing additional content-encodings that have been applied to a *resource* of a DIDL document and thus to indicate which decoding mechanisms need to be applied to the *resource* in order to obtain the MIME media-type identified by the `mimeType` attribute. This is achieved using a `contentEncoding` attribute. This functionality is primarily used to allow a *resource* to be compressed without losing the knowledge regarding its underlying MIME media-type. If multiple content-encodings have been applied to the *resource*, each content-encoding is listed in the value of the `ContentEncoding` attribute in a space-delimited list in the order in which they were applied [3].

In addition to the above attributes, the second edition of MPEG-21 DID allows for `Resource` elements to have attributes from other XML Namespaces (not defined by MPEG-21 DID), enabling the provision of application or community specific information about that *resource*.

## 6.2. Providing secondary information

Secondary textual information, *statements*, are contained inside a `Statement` element. Whereas *resources* are considered identifiable assets, *statements* cannot be identified, and hence, should not be treated as individual assets in their own right. `Statement` elements are embedded inside a `Descriptor` elements, forming so-called *descriptor/statement* constructs.

As indicated in Table 1, a *statement* can be provided By Value and/or By Reference. Just like a `Resource` element, a `Statement` element has a required `mimeType` attribute, and optional `ref`, `encoding`, and `contentEncoding` attributes. Table 4 shows a `Statement` element containing a By Value provision of secondary information about the sample scholarly publication, expressed in the Dublin Core format.

```
<didl:Statement mimeType="text/xml; charset=UTF-8">
  <oai_dc:dc xmlns:oai_dc="http://www.openarchives.org/OAI/2.0/oai_dc/"
    xmlns:dc="http://purl.org/dc/elements/1.1/"
    xmlns:dcterms="http://purl.org/dc/terms/">
    <dc:title>Key Concepts in the Architecture of the Digital Library</dc:title>
    <dc:creator>William Y. Arms</dc:creator>
  </oai_dc:dc>
<didl:Statement>
```

**Table 4**: A `Statement` element containing a 'By Value' provision of the secondary information expressed in the DC format

## 6.3. More about `Descriptor/Statement` constructs

`Descriptor/Statement` elements, representing a *descriptor/statement* construct, provide an extensible mechanism to associate textual secondary information with entities of the Abstract Model. For example, in order to associate – say – an identifier with an *item*, a *descriptor* containing the identifier *statement* can be created as a child element of the `Item` element.

As will be shown, the MPEG-21 framework itself defines ways to use *descriptor/statement* constructs as a means to convey – amongst others – identification information, rights information, and processing information. This approach is illustrated in Section 6.3.1. To facilitate the provision of domain or application-specific information, *descriptor/statement* constructs may also be defined by third parties.

In order to do so, typically, an XML Schema with an associated XML Namespace is created to contain elements and attributes required to address specific needs. This approach is illustrated in Section 6.3.2.

### 6.3.1. MPEG-21 defined `Descriptor/Statement` constructs

#### 6.3.1.1. MPEG-21 DII: Using `Descriptor/Statement` constructs to identify Digital Items

Through the introduction of a special Part, the MPEG-21 Digital Item Identification (MPEG-21 DII), MPEG-21 recognizes the importance of identifiers in network-based applications. MPEG-21 DII specifies the usage of `Descriptor/Statement` constructs for the identification of Digital Items and parts thereof. To that end, it introduces a DII XML Namespace with elements that can be used to associate identifiers and/or types with *container*, *item*, *component*, and *anchor* entities. For example, Table 5 shows the use of the `Identifier` element from the DII XML Namespace to associate the URI 'info:doi/10.1045/july95-arms' with the sample Digital Item; where 'doi:10.1045/july95-arms' is a unique DOI [43] number for the publication, and 'info' is a proposed URI Scheme [51].

```
<didl:Item>
  <didl:Descriptor>
    <didl:Statement mimeType="text/xml; charset=UTF-8">
      <dii:Identifier xmlns:dii="urn:mpeg:mpeg21:2002:01-DII-NS">
      info:doi/10.1045/10.1045/july95-arms</dii:Identifier>
    </didl:Statement>
  </didl:Descriptor>
  ...
</didl:Item>
```

**Table 5**: A `Descriptor/Statement` construct containing an identifier expressed in the DII Namespace

MPEG-21 DII also defines a `RelatedIdentifier` element that carries identifiers that are 'related' to a Digital Item or a part thereof. The specification, however, remains silent regarding the nature of the relationship of the thing identified by a `RelatedIdentifier` to the Digital Item (or constituent thereof) that carries the `RelatedIdentifier`. At the time of writing, the authors are involved in an amendment aimed at qualifying the nature of this relationship by means of a `relationshipType` attribute. The value of this attribute is in the form of a URI and corresponds to a verb from the MPEG-21 Rights Data Dictionary (MPEG-21 RDD) [31].

## 6.3.1.2. MPEG-21 REL: Using `Descriptor/Statement` constructs to associate rights expressions with Digital Items

MPEG-21 Rights Expression Language (MPEG-21 REL) specifies the usage of `Descriptor/Statement` constructs to associate rights expressions with a Digital Item or parts thereof. This is achieved through the introduction of a language inspired by XrML [16] with elements and attributes in a REL XML Namespace. Table 6 shows the use of the `license` element of the REL XML Namespace to associate very basic copyright information with the sample Digital Item. MPEG-21 IPMP, currently under development, will provide tools to enforce rights expressions declared by the REL.

```xml
<didl:Item>
  <didl:Descriptor>
    <didl:Statement mimeType="text/xml; charset=UTF-8">
      <r:license xmlns:r="urn:mpeg:mpeg21:2003:01-REL-R-NS">
        <!-- licenses can be added here -->
        <r:otherInfo>
          <dc:rights xmlns:dc="http://purl.org/dc/elements/1.1/">
          Copyright 1995; Corporation for National Research Initiatives</dc:rights>
        </r:otherInfo>
      </r:license>
    </didl:Statement>
  </didl:Descriptor>
  ...
</didl:Item>
```

**Table 6**: A `Descriptor/Statement` construct containing a simple rights expression

## 6.3.1.3. MPEG-21 DIP: Using `Descriptor/Statement` constructs to associate processing information with Digital Items

MPEG-21 Digital Item Processing (MPEG-21 DIP) specifies an architecture pertaining to the dissemination of Digital Items. This, yet to be standardized, MPEG-21 Part introduces the concept of a Digital Item Method (DIM), as a way to allow a Digital Item author to provide suggested interactions of an end-User with a Digital Item. A DIM is physically contained in the same DIDL document as the Digital Item with which it is associated. In the current practice, a DIM is associated with an element of a DIDL document using a special-purpose `Descriptor/Statement` construct, containing elements of the DIP XML Namespace. The DIMs (and the associated Digital Items parts) can be extracted from a DIDL document and processed upon request of an agent by a special component, called an MPEG-21 DIP Engine.

### *6.3.2. Non-MPEG-21 defined `Descriptor/Statement` constructs*

`Descriptor/Statements` constructs may also be used by third parties to provide information that is specific to the nature of the content, the application and the community. Table 7 shows the use of a `Descriptor/Statement` construct to associate the *statement* represented in Table 4 with an `Item`.

The second edition of MPEG-21 DID also allows for the association of third party information with DIDL elements via XML attributes. For instance, an `Item` element may have attributes from other XML Namespaces providing information about that *item*. Note that, as is the case with the `Descriptor/Statement` element construct, these attributes are considered XML instantiations of the abstract *descriptor/statement* entities.

```xml
<didl:Item>
   <didl:Descriptor>
      <didl:Statement mimeType="text/xml; charset=UTF-8">
         <oai_dc:dc xmlns:oai_dc="http://www.openarchives.org/OAI/2.0/oai_dc/"
            xmlns:dc="http://purl.org/dc/elements/1.1/"
            xmlns:dcterms="http://purl.org/dc/terms/">
            <dc:title>Key Concepts in the Architecture of the Digital Library</dc:title>
            <dc:creator>William Y. Arms</dc:creator>
         </oai_dc:dc>
      <didl:Statement>
   </didl:Descriptor>
   ...
</didl:Item>
```

**Table 7**: Community specific use of the `Descriptor/Statement` construct

## 6.4. Putting the DIDL document together

So far, the `Resource` and `Statement` elements have been introduced, and the usage of `Descriptor/Statement` constructs has been shown to enable associating secondary information with a Digital Item or its constituents. This section uses these building blocks to compile a DIDL document that represents the sample Digital Item. The explanation works its way up from the `Resource` and `Component` elements to the `Item` element, and then all the way up to the `DIDL` root element. The complete DIDL document, representing the sample digital asset, is shown in Table 9.

A `Component` element, representing a *component*, is introduced to bind one or more `Resource` elements to a (set of) `Descriptor` elements. In MPEG-21 DID first edition, the *resources* contained inside a single *component* are considered 'equivalent'. In MPEG-21 DID second edition, based on input from the authors, 'equivalence' has been further qualified to mean 'bit equivalent' [5]. As a result, MPEG-21 DID allows for the provision of multiple bit equivalent *resources* inside a single *component*, using both the By Value and By Reference approaches. This feature is appealing when the same data is available at multiple locations.

Table 8 shows a `Component` element wrapping the PDF *resources* represented in Table 2 and Table 4. The first *resource* is provided By Reference through the inclusion of a reference to its network-location; the second, bit-equivalent, *resource* is provided By Value, and is base64 encoded.

```
<didl:Component>
   <didl:Resource mimeType="application/pdf" ref="http://purl.lanl.gov/tech/pdf/015997845.pdf"/>
   <didl:Resource mimeType="application/pdf" encoding="base64">
   JVBERi0xLjMKJeLjz9MNCjEgMCBvYmoKPDwgCi9TdWJqZWN0ICgpCi9LZXl3b3JkcyAoKQovQ3Jl
   YXRvciAoWFBQKQovVGl0bGUgKCkKL1Byb2R1Y2VyICgpNb2REYXRlIChEOjIwMDQwNTE0MTIy
   MzE2KQovQ3JlYXRpb25EYXRlICgyMDA0MDUxNDEyMjMwMSkKL0F1dGhvciAoKQo+PiAKZW5kb2Jq
   CjIgMCBvYmoKPDwgCi9UeXBlIC9QYWdlIAovUGFyZW50IDExIDAgUiAKL1Jlc291cmNlcyA0IDAg...</didl:Resource>
</didl:Component>
```

**Table 8**: A `Component` element grouping bit equivalent *resources*.

In addition to the PDF *resource* provided in the *component* of Table 8, the sample Digital Item also contains the PostScript *resource*. Because the PostScript *resource* is not bit equivalent with the PDF file, it is provided in a separate *component*. And, because the PostScript *resource* and the PDF *resource* share an identifier, both *components* are provided in the same *item*. This *item* is considered to be the declarative representation of the sample Digital Item. An `Item` element is introduced to contain both *components*, and to bind them to a set of *descriptor/statement* constructs. The `Descriptor/Statement` constructs associated with the `Item` contain secondary information about the Digital Item. A first `Descriptor/Statement` construct conveys the DOI of the Digital Item using the `Identifier` element from the DII XML namespace. A second `Descriptor/Statement` construct conveys secondary information about the Digital Item expressed in the Dublin Core format. The DIDL document itself opens with the `DIDL` root element containing the DID XML Namespace declaration.

```
<?xml version="1.0" encoding="UTF-8"?>
<didl:DIDL xmlns:didl="urn:mpeg:mpeg21:2002:02-DIDL-NS">
   <didl:Item>
```

```xml
        <didl:Descriptor>
          <didl:Statement mimeType="text/xml; charset=UTF-8">
            <dii:Identifier xmlns:dii="urn:mpeg:mpeg21:2002:01-DII-NS">
            info:doi/10.1045/july95-arms</dii:Identifier>
          </didl:Statement>
        </didl:Descriptor>
        <didl:Descriptor>
          <didl:Statement mimeType="text/xml; charset=UTF-8">
            <oai_dc:dc xmlns:oai_dc="http://www.openarchives.org/OAI/2.0/oai_dc/"
              xmlns:dc="http://purl.org/dc/elements/1.1/"
              xmlns:dcterms="http://purl.org/dc/terms/">
              <dc:title>Key Concepts in the Architecture of the Digital Library</dc:title>
              <dc:creator>William Y. Arms</dc:creator>
            </oai_dc:dc>
          <didl:Statement>
        </didl:Descriptor>
        <didl:Component>
          <didl:Resource mimeType="application/pdf" ref="http://purl.lanl.gov/tech/pdf/015997845.pdf"/>
          <didl:Resource mimeType="application/pdf" encoding="base64">
          JVBERi0xLjMKJeLjz9MNCjEgMCBvYmoKPDwgCi9TdWJqZWN0ICgpCi9LZXl3b3JkcyAoKQovQ3Jl
          YXRvciAoWFBBQKQovVGl0bGUgKCkKL1Byb2R1Y2VyICgpCi9Nb2REYXRlIChEOjIwMDQwNTE0MTIy...</didl:Resource>
        </didl:Component>
        <didl:Component>
          <didl:Resource mimeType="application/ps" ref="http://purl.lanl.gov/tech/ps/015997845.ps"/>
        </didl:Component>
    </didl:Item>
  </didl:DIDL>
```

**Table 9**: An `Item` element representing the *item* representing the sample digital asset

In addition, multiple *items* can be grouped in a *container*, represented by a `Container` element (see Table 10). Note that a *container* should not be considered a Digital Item or a digital asset in its own right, but rather a mechanism that allows the grouping of Digital Items in a single document.

```xml
<?xml version="1.0" encoding="UTF-8"?>
<didl:DIDL xmlns:didl="urn:mpeg:mpeg21:2002:02-DIDL-NS">
  <didl:Container>
      <didl:Item>...</didl:Item>
      <didl:Item>...</didl:Item>
      <didl:Item>...</didl:Item>
  </didl:Container>
</didl:DIDL>
```

**Table 10**: A `Container` element grouping various *items*

# 7. Using MPEG-21 DID at the LANL Research Library

So far, this article has provided an insight in the basic concepts of the MPEG-21 DID Abstract Model and MPEG-21 DIDL. This section illustrates the feasibility and attractiveness of MPEG-21 DID for digital library applications by briefly describing its use in the aDORe repository effort [50] of the Research Library of the Los Alamos National Laboratory (LANL).

The LANL Research Library locally stores TeraBytes of digital content from primary and secondary scholarly publishers including Elsevier, American Physical Society, ACM, IEEE, Wiley, etc. A large scale project is ongoing aimed at representing all these scholarly digital assets using MPEG-21 DID. At the time of writing, the LANL aDORe Repository contains 30,000,000 DIDL documents, a figure that is expected to triple in the next 12 months. For each digital asset that is obtained from a scholarly publisher, and that is to be ingested into aDORe, a DIDL document is created that physically contains and/or references the datastreams of which the asset consists. Apart from these primary datastreams, this DIDL document also contains secondary information pertaining to the asset and/or parts thereof. Some of this secondary information is directly taken from the asset as it was obtained from the information provider. Other secondary information is added by the ingestion process and is crucial for the functioning of aDORe and its directly associated processes. This secondary information is provided by means of `Descriptor/Statement` constructs or XML attributes. As will be shown in this section, special purpose `Descriptor/Statement` constructs are used to convey information about a Digital Item and its constituent datastreams, while the `DIDLInfo` child element of the `DIDL` root element, as well as attributes attached to that root element, are used to convey information about the DIDL document itself.

## 7.1. Identifiers

Each digital asset obtained from a publisher is regarded an MPEG-21 Digital Item. Therefore, during the ingestion process, the digital asset is mapped to an `Item` element that contains all information (primary and secondary) related to the asset. A `Descriptor/Statement` construct attached to the `Item` is used to convey the identifier of the digital asset; this is done by means of the `Identifier` element of the MPEG-21 DII XML Namespace. This identifier corresponds to what the OAIS Reference Model [25] categorizes as Content Information Identifiers. As shown in the sample DIDL document in Section 7.5, the Content Information Identifier of the sample Digital Item is 'info:doi/10.1045/july95-arms'

The aforementioned `Item` is part of a DIDL document, which functions as an OAIS Archival Information Package (OAIS AIP) in aDORe. At the time of ingestion, each such DIDL document is accorded a globally unique OAIS Archival Information Package Identifier. These Package Identifiers are constructed using the UUID algorithm [39], and are expressed as URIs in the 'info:lanl-repo/' namespace, which the LANL Research Library has registered under the info URI Scheme [51]. The OAIS Package Identifier of each AIP is conveyed as the value of the `DIDLDocumentId` attribute of the DIDL root element. The value of the OAIS Package Identifier of the sample DIDL document is 'info:lanl-repo/i/00002cb8-c477-11d8-a819-b1db893d21e6'. In addition to the OAIS Package Identifier pertaining to the DIDL document, all `Items` and `Components` of which the DIDL document consists are accorded globally unique XML IDs, again created using the UUID algorithm.

Resulting from the described approach, to accord and convey identifiers, are two parallel URI mechanisms to address assets stored in aDORe. Figure 4 illustrates these mechanism for the sample Digital Item. One mechanism is directly related to the OAIS AIP notion, and is based on XML technology. It uses the OAIS Package Identifier conveyed in the `DIDLDocumentId` attribute of the `DIDL` root element to identify DIDL documents, and this same Package Identifier extended with an XML ID fragment to identify XML child elements of the DIDL documents. The other mechanism is directly related to the OAIS Content Information notion. It uses the OAIS Content Information Identifiers of the digital assets provided in the `Identifier` element of the MPEG-21 DII XML Namespace, which is typically attached using a `Descriptor/Statement` construct to the `Item` that represents the Digital Item. Content Information Identifiers can be provided at levels below this `Item` too, for example, if individual datastreams of which a Digital Item consists have identifiers of their own.

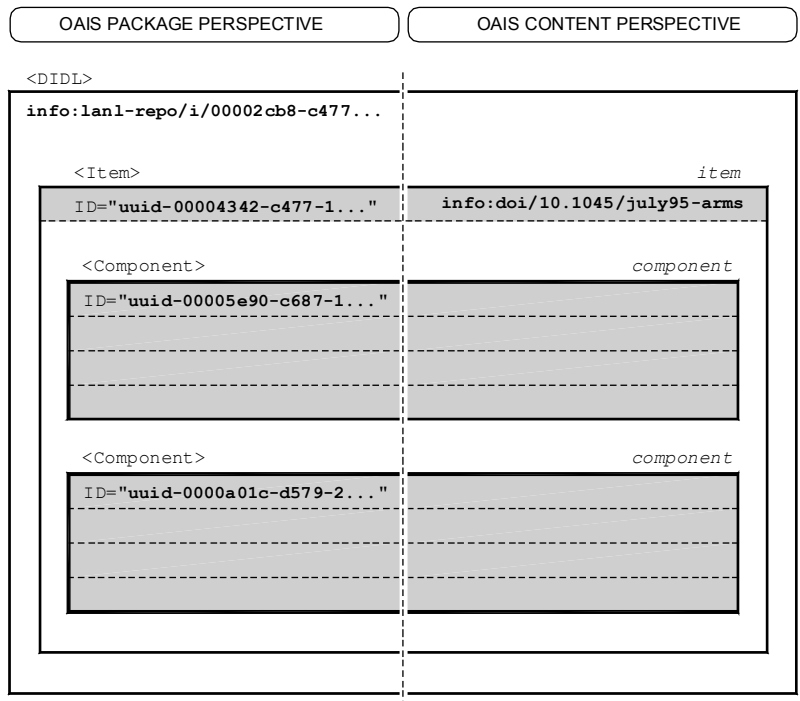

**Figure 4**: Two complete and parallel identification mechanisms

Both types of OAIS Identifiers are used in the protocols through which assets stored in aDORe are accessible:

- The Open Archives Initiative Protocol for Metadata Harvesting (OAI-PMH) [38] is used as the protocol to access DIDL documents. This allows for batch access (using OAI-PMH `ListRecords`) to DIDL documents based on time (range) of their ingestion, and access to an individual DIDL document (using OAI-PMH `GetRecord`) by means of its OAIS Package Identifier. The response to the OAI-PMH GetRecord request corresponds to the OAIS concept of Dissemination Information Package (OAIS DIP).
- NISO OpenURL [45] is used as the technique to request context-sensitive services pertaining to stored Digital Items. The crucial key on these OpenURLs conveys the OAIS Content Information Identifier of the Digital Item or constituent for which a context-sensitive service is requested.

These are appealing illustrations of how OAIS Identifier concepts can be implemented using MPEG-21 DID. In the context of aDORe, the parallel addressing mechanism is also attractive for practical purposes. Consider that an updated version of a previously ingested digital asset is obtained from a publisher. In this case, both the existing and the newly obtained digital asset will share a Content

Information Identifier. Through the ingestion process, however, both assets will be contained in different DIDL documents, each of which has its own, globally unique, OAIS AIP Identifier. As a result, the OAIS Content Information Identifier is used to address all versions and copies of a Digital Item, whereas the OAIS Package Identifier allows to distinguish between versions and copies of the same Digital Item.

## 7.2. Relationships

Special-purpose `Descriptor/Statement` constructs are introduced to express relationships within and among Digital Items. These `Descriptor/Statement` constructs contain RDF *statements* expressing relationships such as 'isMemberOf' and 'isTranslationOf'. The RDF *statements* are based on an experimental ontology of relationships that will be evolved towards interoperability in collaboration with colleagues from Cornell University, in the context of the NSF-funded Pathways project.

It should be noted that the relationships provided in these `Descriptor/Statement` constructs are different from the structural relationships that are inherent to the structure of the DIDL document itself. Relationships, such as 'hasResource', 'isPartOfItem', and 'hasIdentifier' can be dynamically derived from the structure of the DIDL document itself, and hence, do not necessarily need to be explicitly included as RDF statements within the DIDL document.

## 7.3. Creation datetimes and formats

Different creation times are associated with digital assets that are ingested in the LANL Repository:

- The creation date of a DIDL document is conveyed by means of a `DIDLDocumentCreated` attribute that is attached to the `DIDL` root element. This attribute is from an XML Namespace defined as part of the aDORe effort. The value of this attribute is used as the datestamp for OAI-PMH access to aDORe. The creation datetime of the sample DIDL document shown in Section 7.5 is '2004-11-22T18:07:18Z'.
- When known, the creation datetime of a datastream of a Digital Item is conveyed using a `Descriptor/Statement` construct attached to the `Component` that contains the datastream as a *resource*. This construct conveys the creation time using the `created` element from the Dublin Core Element Set [44], which was imported into an aDORe XML Namespace. The creation dates of the PDF and Postscript files of the sample Digital Item are '2003-10-29T18:07:18Z' and '2003-10-26T10:03:12Z', respectively.

Also, each `Component` element may have a `Descriptor/Statement` construct that conveys a unique format identifier, capturing fine-grained media format information, of the constituent datastream contained inside the `Component`. The sample document in Section 7.5 show the use of `dc:format` elements to convey format identifiers of the PDF and PostScript datastreams of the sample asset. At the time of writing, investigations are ongoing aimed at expressing these formats by elements from the PREMIS effort [http://www.oclc.org/research/projects/pmwg/]. Also, the values of these elements could be expressed using controlled format registries, such as the Global Digital Format Registry (GDFR) [http://hul.harvard.edu/gdfr/] and the PRONOM File Format registry [http://www.nationalarchives.gov.uk/pronom/].

## 7.4. Digests and signatures

W3C XML Signatures [2] embedded in DIDL documents are used to allow verifying the accuracy and authenticity of assets stored in aDORe:

- To allow verifying the DIDL document itself, a `Signature` element from the W3C Signature XML Namespace is provided as a child of the `DIDLInfo` element, which itself is a child of the `DIDL` root element. This is illustrated in an abbreviated form in Section 7.5.
- To allow verifying a datastream, a `Signature` element from the W3C Signature XML Namespace is provided in a `Descriptor/Statement` construct attached to the `Component` that contains the datastream as a *resource*. Again, this is illustrated in an abbreviated form in Section 7.5.

W3C XML Signatures are also used in conjunction with DIDL documents in a collaborative project between the American Physical Society (APS) and the LANL Research Library. In this project, aimed at permanently mirroring the collection of the APS at LANL, the OAI-PMH is used as a protocol to recurrently harvest assets represented as DIDL documents from the APS. In order to allow for the verification of the accurate transfer of the assets, XML Signatures are included in the exposed DIDL documents [11].

## 7.5. Sample aDORe DIDL document

```
<?xml version="1.0" encoding="UTF-8"?>
<didl:DIDL DIDLDocumentId="info:lanl-repo/i/00002cb8-c477-11d8-a819-b1db893d21e6"
    diext:DIDLDocumentCreated="2004-11-22T18:07:18Z"
    xmlns:didl="urn:mpeg:mpeg21:2002:02-DIDL-NS"
    xmlns:diext="http://library.lanl.gov/2005-08/aDORe/DIDLextension/">
    <didl:DIDLInfo>
        <dsig:Signature xmlns:dsig="http://www.w3.org/2000/09/xmldsig#">...</dsig:Signature>
```

```xml
      </didl:DIDLInfo>
      <didl:Item id="uuid-00004342-c477-11d8-a819-b1db893d21e6">
         <didl:Descriptor>
            <didl:Statement mimeType="text/xml; charset=UTF-8">
               <dii:Identifier xmlns:dii="urn:mpeg:mpeg21:2002:01-DII-NS">
               info:doi/10.1045/july95-arms</dii:Identifier>
            </didl:Statement>
         </didl:Descriptor>
         <didl:Descriptor>
            <didl:Statement mimeType="text/xml; charset=UTF-8">
               <oai_dc:dc xmlns:oai_dc="http://www.openarchives.org/OAI/2.0/oai_dc/"
                  xmlns:dc="http://purl.org/dc/elements/1.1/"
                  xmlns:dcterms="http://purl.org/dc/terms/">
                  <dc:title>Key Concepts in the Architecture of the Digital Library</dc:title>
                  <dc:creator>William Y. Arms</dc:creator>
               </oai_dc:dc>
            <didl:Statement>
         </didl:Descriptor>
         <didl:Component id="uuid-00005e90-c687-11d8-a819-b1db893d21e6">
            <didl:Descriptor>
               <didl:Statement mimeType="text/xml; charset=UTF-8">
                  <diadm:Admin xmlns:diadm="http://library.lanl.gov/2004-01/STB-RL/DIADM">
                     <dc:format xmlns:dc="http://purl.org/dc/elements/1.1/">info:lanl-repo/fmt/5</dc:format>
                     <dcterms:created xmlns:dcterms="http://purl.org/dc/terms/">
                        2003-10-29T18:07:18Z</dcterms:created>
                  </diadm:Admin>
               </didl:Statement>
            </didl:Descriptor>
            <didl:Descriptor>
               <didl:Statement mimeType="text/xml; charset=UTF-8">
                  <dsig:Signature xmlns:dsig="http://www.w3.org/2000/09/xmldsig#">...</dsig:Signature>
               </didl:Statement>
            </didl:Descriptor>
            <didl:Resource mimeType="application/pdf" ref="http://purl.lanl.gov/tech/pdf/015997845.pdf"/>
            <didl:Resource mimeType="application/pdf" encoding="base64">
            JVBERi0xLjMKJeLjz9MNCjEgMCBvYmoKPDwgCi9TdWJqZWN0ICgpCi9LZXl3b3JkcyAoKQovQ3Jl
            YXRvciAoWFBQKQovVGl0bGUgKCkKL1Byb2R1Y2VyICgpL9Nb2REYXRlIChEOjIwMDQwNTE0MTIy...</didl:Resource>
         </didl:Component>
         <didl:Component id="uuid-0000a01c-d579-21d8-a819-b1db893d21e6">
            <didl:Descriptor>
               <didl:Statement mimeType="text/xml; charset=UTF-8">
                  <diadm:Admin xmlns:diadm="http://library.lanl.gov/2004-01/STB-RL/DIADM">
                     <dc:format xmlns:dc="http://purl.org/dc/elements/1.1/">info:lanl-repo/fmt/10</dc:format>
                     <dcterms:created xmlns:dcterms="http://purl.org/dc/terms/">
                        2003-10-26T10:03:12Z</dcterms:created>
                  </diadm:Admin>
               </didl:Statement>
            </didl:Descriptor>
            <didl:Descriptor>
               <didl:Statement mimeType="text/xml; charset=UTF-8">
                  <dsig:Signature xmlns:dsig="http://www.w3.org/2000/09/xmldsig#">...</dsig:Signature>
               </didl:Statement>
            </didl:Descriptor>
            <didl:Resource mimeType="application/ps" ref="http://purl.lanl.gov/tech/ps/015997845.ps"/>
         </didl:Component>
      </didl:Item>
```

```
    </didl:DIDL>
```

## 8. Conclusion

This article has described the MPEG-21 Digital Item Declaration. It has also illustrated its use in a specific digital library context. Hopefully, this has resulted in an insight that embracing MPEG-21 DID for digital library applications is both feasible and attractive.

From a strategic perspective, MPEG-21 DID is appealing because it is part of the MPEG suite of ISO standards which is likely to receive strong industry backing [14]. Also, MPEG-21 DID is part of a broader architecture, many components of which are of direct relevance to the digital library community. And, importantly, MPEG-21 DID is an ISO standard developed by major players in the content and technology industry, which provides some guarantees regarding its adoption and the emergence of compliant tools.

From a functional perspective, MPEG-21 DID is attractive because of the existence of a well-specified Abstract Model, and the capability this yields for the representation of Digital Items in various evolving technological environments while maintaining the basic architectural concepts. Also, MPEG-21 DID takes an approach that enforces cross-community interoperability, while allowing the flexibility for the emergence of compliant, community specific profiles. The XML Schema of the MPEG-21 DIDL instantiation is elegant and simple, making the development of tools quite straightforward. Especially appealing is the flexibility and extensibility provided by the *descriptor/statement* approach, instantiated in MPEG-21 DIDL by means of the `Descriptor`/`Statement` elements and XML attributes. The MPEG-21 standard itself makes use of those *descriptor/statement* constructs to provide a fundamental solution for the identification of Digital Items, to associate processing methods with Digital Items, and to express rights related to Digital Items. *Descriptor/statement* constructs can also be used to address community-specific interoperability requirements.

Through the brief description of the use of MPEG-21 DID at the LANL Research Library, it was shown that implementation in a manner consistent with fundamental OAIS concepts is feasible. It was also shown that the MPEG-21 DID technologies can be integrated with technologies such as OAI-PMH, OpenURL, and W3C XML Signatures that have emerged from other information communities. Interested readers are referred to [4,8,34,50] for a more in-depth overview of the LANL aDORe architecture, and the role MPEG-21 and other technologies play in it.

## Acknowledgements

The authors would like to thank their colleagues Lyudmila Balakireva, Xiaoming Liu, and Thorsten Schwander of the LANL Digital Library Research and Prototyping Team, as well as Patrick Hochstenbach (previously at LANL, and now at Ghent University) and Henry Jerez (previously at LANL, and now at CNRI) for their contributions to the reported work. Many thanks also to Rick Luce, the LANL library director, for his ongoing support for the reported work. Jeroen Bekaert also wishes to thank the Fund for Scientific Research (Flanders, Belgium) for his Ph.D. scholarship. The reported work is partially funded by a grant from the Library of Congress's National Digital Information Infrastructure Program.